\documentstyle[12pt,a4,epsf]{article}
\global\arraycolsep=2pt 

\begin{document}
\makeatletter
\def\fmslash{\@ifnextchar[{\fmsl@sh}{\fmsl@sh[0mu]}}
\def\fmsl@sh[#1]#2{%
  \mathchoice
    {\@fmsl@sh\displaystyle{#1}{#2}}%
    {\@fmsl@sh\textstyle{#1}{#2}}%
    {\@fmsl@sh\scriptstyle{#1}{#2}}%
    {\@fmsl@sh\scriptscriptstyle{#1}{#2}}}
\def\@fmsl@sh#1#2#3{\m@th\ooalign{$\hfil#1\mkern#2/\hfil$\crcr$#1#3$}}
\makeatother
\def\g5{\gamma_5}
\def\ml{m_\Lambda}
\def\mb{m_{\Lambda_b}}
\def\beq{\begin{equation}}
\def\eeq{\end{equation}}
\def\beqa{\begin{eqnarray}}
\def\eeqa{\end{eqnarray}}
\def\ps{{\rm{\bf\hat{p}.S}_\Lambda}}
%
\thispagestyle{empty}
\begin{titlepage}

\begin{flushright}
hep-ph/9701399 \\
TTP97--06 \\
January 28, 1997
\end{flushright}

\vspace{0.3cm}
\boldmath
\begin{center}
\Large\bf Flavour Changing Neutral Current Decays of Heavy Baryons: \
          The Case $\Lambda_b \to \Lambda \gamma$ 
\end{center}
\unboldmath
\vspace{0.8cm}

\begin{center}
{\large Thomas Mannel} and {\large Stefan Recksiegel} \\
{\sl Institut f\"{u}r Theoretische Teilchenphysik,
     D -- 76128 Karlsruhe, Germany.} 
\end{center}

\vspace{\fill}

\begin{abstract}
\noindent
We investigate the rare decay $\Lambda_b \to \Lambda \gamma$ which receives
both short and long distance contributions. We estimate the long distance 
contributions and find them very small. The form factors are obtained 
from $\Lambda_c \to \Lambda \ell \bar{\nu}_\ell$ using heavy quark symmetry
and a pole model. The short distance piece opens a window to new physics 
and we discuss the sensitivity of  $\Lambda_b \to \Lambda \gamma$ 
to such effects. 
\end{abstract}
\end{titlepage}


\addtocounter{page}{1}
\section{Introduction}
Flavour Changing Neutral Current (FCNC) processes have attracted
renewed attention since the recent CLEO measurement of the FCNC decays
of the type $b \to s \gamma$. In the Standard Model (SM) these
processes are forbidden at the tree level and are strongly suppressed
by the GIM mechanism, in particular for up type quarks. Hence they
offer a unique possibility to test the CKM sector of the SM and
possibly open a window to physics beyond the SM.

From the experimental side there is some data on the exclusive decay
$B \to K^* \gamma$ and the inclusive decay $B \to X_s \gamma$ with
branching fractions \cite{pdg}
\begin{eqnarray*}
Br (B^+ \to K^{+*} \gamma) &=& (5.7 \pm 3.3) \times 10^{-5} \\
Br (B^0 \to K^{0*} \gamma) &=& (4.0 \pm 1.9) \times 10^{-5} \\
Br (B \to X_s \gamma) &=& (2.3 \pm 0.7) \times 10^{-4}
\end{eqnarray*}
which is at present compatible with the SM. However, the experimental as
well as the theoretical uncertainties are still sizable; in
particular an improvement of the theoretical prediction involves
a next-to-leading order calculation which has recently become 
available \cite{chet}.

Various scenarios have been used to calculate the influence of
physics beyond the SM on the decay $b \to s \gamma$ and also
model independent analyses have been suggested \cite{joanne}.
However, based
on the decays of $B$ mesons it will not be possible to analyse
the helicity structure of the effective hamiltonian mediating the
decay $b \to s \gamma$, since the information on the handedness of the
quarks is lost in the hadronization process.

The only way to access the helicity of the quarks is to consider
the decay of baryons. From the experimental side the decay
$\Lambda_b \to \Lambda \gamma$ is a good candidate, since the
subsequent $\Lambda$ decay $\Lambda \to p \pi$ is self analyzing.
The only experimental drawback of this mode is that the production
rate of $\Lambda_b$ baryons in $b$ quark hadrionization  
is about an order of magnitude smaller
than the rate for $B$ mesons, and hence the analysis suggested here
has to wait for more data on heavy quark decays from colliders.

From the theoretical point of view one may make use of the heavy
quark symmetry for the $b$ quark. The $b \to s$ transition is
of the heavy-to-light type for which heavy quark symmetries restrict
the number of form factors for the baryonic transition 
$\Lambda_Q \to $ light spin-1/2 baryon to only two.
These heavy quark symmetries are expected to work best at the point
of maximal momentum transfer $q=q_{max}$ (corresponding to small recoil 
to the light degrees of freedom), 
whereas in the decay under consideration we are at the opposite
side of phase space, namely at $q^2 = 0$. We shall still use heavy
quark symmetry, at least it will be a reasonable model assumption for
the form factors. Once we use heavy quark symmetries over the whole 
phase space, one may obtain the form factors from the semileptonic decay
$\Lambda_c \to \Lambda \ell \bar{\nu}_\ell$, for which some data exists
\cite{CLEOLamc}. Unfortunately, data are still not good enough to allow
a measurement of the two form factors as a function of $q^2$, and consequently 
one has to model the form factor shape using some ansatz. This makes the 
predicted total rate to some extent uncertain, while the polarization of 
the $\Lambda$ is practically unaffected by the choice of a model.  

Another theoretical difficulty with the process suggested are possible
long distance contributions which will dilute the effects of a
non-SM-contribution to the short distance effective hamiltonian 
for $b\to s \gamma$.
There are two effects of this kind: A vector dominance contribution
from the process $b \to s (\bar{c}c) \to s J/\Psi \to s \gamma$
and the internal $W$ exchange. Both contributions are hard to estimate,
since one has to refer to models.

In the next section we shall discuss the short distance contribution
and its hadronic matrix elements. In section 3 we shall use simple
models to estimate the long distance contributions. In section 4 we shall 
use the input of the semileptonic $\Lambda_c$ decay and two simple models 
for the form factors to obtain a prediction for the total rate of 
$\Lambda_b \to \Lambda \gamma$. Section 5 is devoted to a discussion 
of the polarization variable and its connection to the parameters 
of the short distance part of the effective hamiltonian. The standard model 
and the sensitivity to non-standard model effects are considered. Finally we
discuss our results and conclude.

\section{The short distance contribution}
In the SM as well as in many non-SM scenarios, the hamiltonian 
relevant for $b \to s$ transitions consists
of ten operators ${\cal O}_i, i=1,...10$, of which only one
(${\cal O}_7$) mediates the decay $b \to s \gamma$. The effective
hamiltonian has the form
\begin{equation}
H_{eff} = \frac{4 G_F}{\sqrt{2}} V_{ts}^* V_{tb} C_7 (\mu)
          {\cal O}_7 (\mu)
\end{equation}
with
\begin{equation}
{\cal O}_7 = \frac{e}{32 \pi^2} \left[ m_b \bar{s} \sigma_{\mu \nu}
             (1+\gamma_5) b
           + m_s \bar{s} \sigma_{\mu \nu}
             (1-\gamma_5) b \right] F^{\mu \nu}
\end{equation}
where the renormalization scale $\mu$ is usually taken to be $m_b$.
It is well known that the leading-log calculation exhibits a strong residual 
dependence on the renormalization point $\mu$, since the QCD corrections 
encoded in $C_7$ are sizable. Recently the complete next-to-leading order 
calculation has been performed and we shall use the value for $C_7$ from this 
calculation.   

A more general form of the Operator ${\cal O}_7$ allowing for non--SM
couplings is 
\beq
\widehat{\cal O}_7 = \frac{e}{32 \pi^2} m_b \bar{s} \sigma_{\mu \nu}
             (g_V-g_A\gamma_5) b F^{\mu \nu}
\eeq
where $g_V$ and $g_A$ are the vector and axial vector couplings,
respectively. In the Standard Model
\beq
g_V=1+{m_s\over m_b},\quad g_A=-1+{m_s\over m_b}
\eeq
The correction to the integer values of $g_A$ and $g_V$ 
due to the non-vanishing $s$ quark mass is of the order of 3\%.

Hence in general at the level of dimension-six operators there are two 
operators relevant for the decay under discussion, namely the one with 
right-handed $b$ quark (implying left-handed $s$ quark) and the one with 
right-handed $s$ quark (implying left-handed $b$ quark). In other words, in  
general there  are two parameters, which may be taken to be $C_7 g_V$ and 
$C_7 g_A$.

Once the short distance structure of the effective Hamiltonian is known, 
it remains to calculate matrix elements of the operators. In the present case
we shall make use of the fact that the $b$ quark is heavy, while the $s$ 
quark is taken to be a light quark. In general, heavy quark symmetries  
restrict the number of possible form factors quite significantly; in the 
present case there are only two, parametrized by
\cite{baryonsinhqet}
\beq \label{heavytolightdecay}
\left< \Lambda(p,s)|\bar s \Gamma b | \Lambda_b(v,s') \right>
= \bar u_\Lambda (p,s) \{ F_1(p.v)+\fmslash v F_2(p.v)\} \Gamma u_{\Lambda_b}
(v, s') \eeq
where $v$ is the velocity of the heavy $\Lambda_b$--baryon which in HQET is
equal to the velocity of the heavy $b$--quark, $s'$ is the spin of the heavy 
$\Lambda_b$, which due to heavy quark spin symmetry equals the spin of the 
$b$ quark,  
$p$ is the momentum of the $\Lambda$--baryon and $s$ its spin.
Furthermore, $\Gamma$ is an arbitrary Dirac matrix, such that any 
transition between a $\Lambda_Q$ ($Q = b,c$) and a light spin $1/2$ baryon 
is given by the same form factors. In particular, this allows us to compare 
the exclusive semileptonic decay of a $\Lambda_c$ with the process under 
consideration here, although the helicity structure of the relevant 
currents is 
completely different.  

Heavy quark symmetries are expected to work best at the point where 
the final state light hadron is slow (in the rest frame of the decaying 
hadron), such that the momentum transfer to the light degrees of freedom
is small. At the other end of phase space, $q^2 = 0$, it is easy to see 
that the the momentum transfer to the light degrees of freedom scales with the 
mass of the heavy quark, it is not clear whether heavy quark symmetries 
are applicable at this point. However, we shall still make use of heavy quark 
symmetries even for $q^2=0$, taking this in the worst case as a model 
assumption. Keeping this in mind, the matrix element of the operator 
$O_7$ between the initial and final state is
\beqa&& \langle \Lambda(p,s), \gamma(k,\varepsilon)| O_7 | \Lambda_b(v,s')
\rangle \\
&& \qquad = {e\over 16\pi^2} m_b \langle \Lambda(p,s) |\bar s \sigma_{\mu\nu}
{1\over 2} (g_V-g_A\g5) b | \Lambda_b(v,s') \rangle
\langle \gamma(k,\varepsilon)|F^{\mu\nu}|0 \rangle \nonumber \eeqa
The quantity of interest is the decay rate of unpolarized $\Lambda_b$ 
baryons into $\Lambda$ baryons with a definite spin directions $s$.
To obtain from this expression this rate as a function of the 
form factors taken at $q^2 = 0$ is a matter of algebra, and we obtain 
\beqa
\Gamma & = &
{ C_7^2 G_F^2 \over \pi} (V_{ts}^*V_{tb})^2
({e\over 16\pi^2})^2 m_b^2 \mb^3 (1- x^2) \nonumber \\
&& \{ {g_V^2 + g_A^2 \over 2} [
    (1-2x^2+x^4 )|F_1|^2 \nonumber \\
&&\quad + (x-2x^3+x^5)(F_1 F_2^* + F_1^* F_2) \nonumber \\
&&\quad + (x^2-2x^4+x^6)|F_2|^2 ] \nonumber \\
&& +g_V g_A (v.s) [ (2x-2x^3) |F_1|^2 \nonumber \\
&&\quad + (2x^2-2x^4)(F_1 F_2^* + F_1^* F_2) \nonumber \\
&&\quad + (2x^3-2x^5)|F_2|^2 ] \}
\label{gammashortdistance} \eeqa
where $x= \ml/\mb$.

As expected, the terms which depend on the spin of the $\Lambda$ 
are all proportional to $|C_7|^2 g_V g_A$, while the rest of the rate has 
the factor $|C_7|^2 (g_V^2 + g_A^2) / 2$. Thus a polarization analysis of the 
final state $\Lambda$ allows a determination of the ratio $g_A / g_V$. 

In order to comply with the standard definitions we rewrite the rate in terms 
of the polarization variables as defined in \cite{pdg} 
\beq
\Gamma = \Gamma_0 \cdot [ 1 + \alpha' \ps]
\eeq
where ${\rm\bf\hat{p}}$ is the momentum vector of the $\Lambda$
and ${\rm{\bf S}_\Lambda}$ is its spin vector. Using expression 
(\ref{gammashortdistance}) one finds for $\alpha'$
\beqa
\alpha' &=& {(1-x^2)\over (1+x^2)}{(2x-6x^3)+2(2x^2-6x^4)R+
 2x^3 R^2 \over (1-2x^2+x^4)+2(x-2x^3)R_(x^2-2x^4)R^2}
 {2g_V g_A \over g_V^2 + g_A^2}
\\
&=& 0.378 \cdot {2g_V g_A \over g_V^2 + g_A^2} 
\label{alphadef} \eeqa
where we have only used the central value of the 
CLEO measurement \cite{cleo}
\beq
R={F_2 \over F_1}= -0.25 \pm 0.14 \pm 0.08
\eeq
and the ratio of the baryon masses $x=0.20$.

Note that $\alpha '$ vanishes in the limit $\ml \to 0$; the leading order 
is given by 
\beq
\alpha' = 2x {2g_V g_A \over g_V^2 + g_A^2}
\eeq
which for $x = 0.2$ already is a reasonable approximation to the 
value obtained from (\ref{alphadef}).

\section{Long Distance Contributions}
In order to obtain a realistic estimate of the branching ratio and of the
sensitivity on the coupling constants $g_V$ and $g_A$ one also has to 
take into account the long distance contributions. One may distinguish 
between two types of contributions: the vector-meson-dominance 
like contributions, where there is a vector meson intermediate state 
and the internal $W$-boson  
exchange, which corresponds to the weak transition of the heavy $\Lambda_b$ 
into excited $\Lambda$ states decaying subsequently into $\Lambda$ and a photon.   

The vector-dominance like contributions are dominated by the CKM allowed 
contributions where one has a $J/\Psi$ as the internal vector meson. 
They are governed by a different part of the $\Delta B = 1$ hamiltonian 
given by
\beq H_{\rm eff}^{c\bar{c}} = {G_F \over \sqrt{2}} 
V_{cs}^*V_{cb}(C_1 {\cal O}_1 + C_2 {\cal O}_2)
\label{heffshortandcc} \eeq
with
\beqa
{\cal O}_1 &=& (\bar s_\alpha \gamma_\mu(1-\g5) b_\alpha)(\bar c_\beta
  \gamma^\mu(1-\g5)c_\beta),\\
{\cal O}_2&=&(\bar s_\alpha \gamma_\mu(1-\g5) b_\beta)(\bar c_\beta
  \gamma^\mu(1-\g5)c_\alpha),
\eeqa
where $\alpha$ and $\beta$ are $SU(3)_{color}$ indices and $C_1$ and $C_2$ 
are the corresponding Wilson coefficients. 

The decays $\Lambda_b \to \Lambda \Psi ^{(n)}$ ($\Psi^{(n)}$ being the excited
states with the quantum numbers of the $J/\Psi$) 
are exclusive non-leptonic decays
and the result for this part will be strongly dependent on model assumptions. 
Following the usual folklore we keep from ${\cal O}_2$ only the part in which
$c$ and $\bar{c}$ are in a color singlet state and perform factorization with 
this part. Thus we model $\Lambda_b \to \Lambda \Psi^{(n)}$ 
by 
\beq \langle \Lambda \Psi^{(n)} | H_{\rm eff}^{c\bar{c}} 
                                     | \Lambda_b \rangle  
     = {G_F \over \sqrt{2}} 
     V_{cs}^*V_{cb}\left(C_1 + \frac{C_2}{N_c}\right) 
     \langle \Lambda | \bar s \gamma_\mu(1-\g5) b 
                          | \Lambda_b \rangle
     \langle \Psi^{(n)} | (\bar c \gamma^\mu(1-\g5)c 
                          | 0 \rangle
\eeq
It is well known that the combination of Wilson coefficients $C_1 +C_2 / N_c$
($N_c = 3$) comes out very small due to an accidental cancellation between the 
two contributions, yielding a too small rate for e.g. the decay 
$B \to K J/\Psi$. Hence it is more realistic to use for this combination 
the value implied by the data and to put $a_2 =  C_1 +C_2 / N_c = 0.24 \pm 0.04$
\cite{deshpande}. 

Furthermore, we consider a process in which we are dealing with a virtual 
$\Psi^{(n)}$ which is quite far off shell. It has been suggested 
\cite{deshpande} to take into account this effect by assuming a $q^2$ dependent
decay constant for the $\Psi^{(n)}$ according to 
\beq
\langle \Psi^{(n)} (q, \epsilon)  | \bar c \gamma^\mu(1-\g5)c 
| 0 \rangle = i g_\Psi^{(n)} (q^2) \epsilon^*_\mu (q)
\eeq
From the comparison between the leptonic width of the $J/\Psi$ (yielding 
$g_\Psi (m_\Psi^2)$) with the data from $\Psi$-photo-production (yielding 
$g_\Psi (0)$) one obtains the ratio 
\beq
\kappa_\Psi = \frac{g_\Psi (0)}{g_\Psi (m_\Psi^2)} = 0.12 \pm 0.04 
\eeq
corresponding to a suppression by roughly a factor of ten. Unfortunately, 
there is no data for excited $\Psi$'s, so we shall simply assume the same 
suppression factor $\kappa$ for all of them. 

Assembling all the pieces, summing over all $\Psi$ resonances with the 
appropriate quantum numbers and employing the Gordon identity to isolate the 
transverse part one finds  
\beqa \langle \Lambda \gamma | H_{\rm eff}^{c\bar{c}} 
                                     | \Lambda_b \rangle  
     = - {1\over 3} a_2 \kappa \sum_n {g_{\Psi^{(n)}}^2(m_{\Psi^{(n)}}^2) 
     \over m_{\Psi^{(n)}}^2 m_b}
    \langle \Lambda \gamma | \bar s \sigma_{\mu\nu} (1 + \g5) b 
    | \Lambda_b \rangle 
    \langle \gamma |F^{\mu\nu}| 0 \rangle .
\label {heffshortandlong} \eeqa
which takes the same form as the short distance contribution.

\begin{figure}[ht]
\vspace{-.5cm}
 \begin{center}
  \leavevmode
  \epsfxsize=5cm
  \epsffile[169 157 405 325]{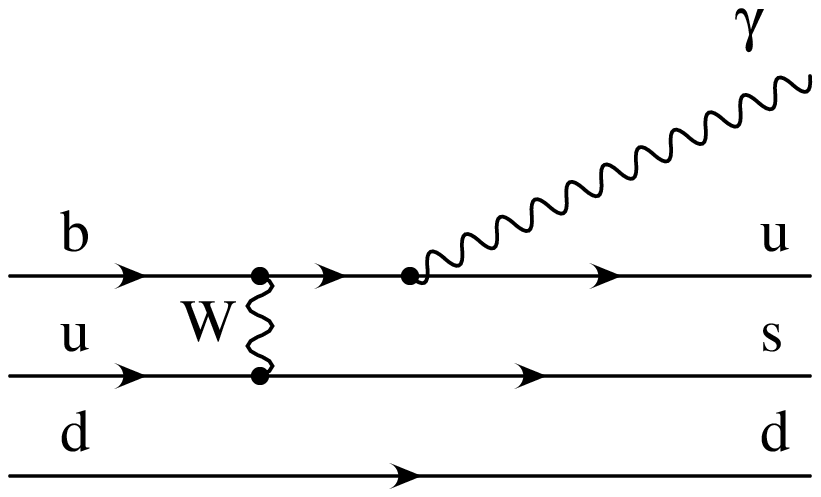}
  \qquad
  \leavevmode
  \epsfxsize=5cm
  \epsffile[169 157 405 325]{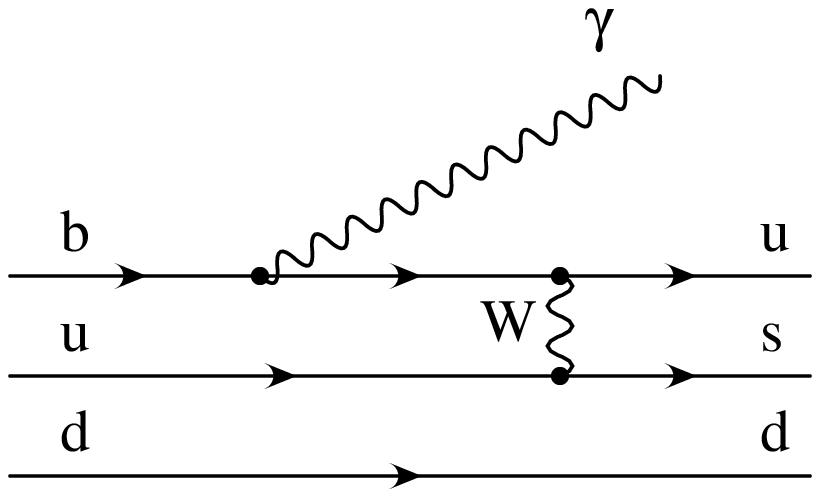}
 \end{center}
\vspace*{-1cm}
\caption{Internal $W$ exchange
\label{internalWex}} \end{figure}

The second kind of long distance contribution is the one originating 
from internal $W$ exchange. The relevant Feynman diagrams are depicted in 
fig.\ref{internalWex}. 
This contribution is even more difficult to 
calculate and we shall perform a relatively crude estimate. The idea we
are going to employ is along the lines of Brodsky and Lepage 
\cite{BrodskyLepage}. In this picture the decay amplitude is decomposed
into a hard piece which is calculated in terms of perturbative QCD while 
the soft contribution is described in terms of wave functions. 

Since in this picture a baryon is still a relatively complicated object, 
we shall simplify further by gathering the up and the down quark in the 
$\Lambda_b$ and the down and strange quark of the $\Lambda$ into diquarks. 
This picture is in fact to some extent
motivated from the phenomenological side, since it has been advertised as
an explanation of the $\Delta I = 1/2$ rule in hyperon decays 
\cite{StechDiQuark}. 

Hence we shall describe the baryons effectively as a bound state of a 
diquark and a $b$ quark or an $u$ quark, respectively. The hard process
in the sense of Brodsky and Lepage is a $b \to u$ transition in which the 
weak boson is absorbed by the $(ud)$ diquark, transforming it into a 
$(sd)$ diquark. In order to conserve energy and momentum a photon is 
radiated off, making this in total a $\Lambda_b \to \Lambda \gamma$ 
transition. 

\begin{figure}[ht]
\vspace{-.3cm}
 \begin{center}
  \leavevmode
  \epsfxsize=2.12cm
  \epsffile[240 145 340 255]{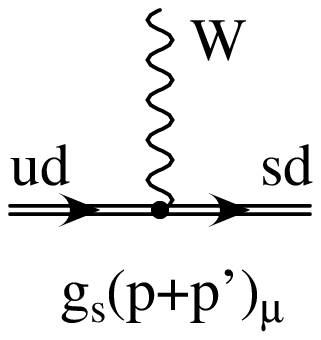}
  \quad
  \epsfxsize=2.12cm
  \epsffile[240 145 340 255]{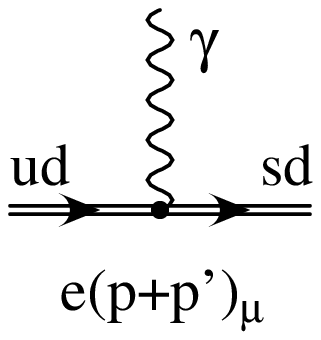}
  \quad
  \epsfxsize=2.12cm
  \epsffile[240 145 340 255]{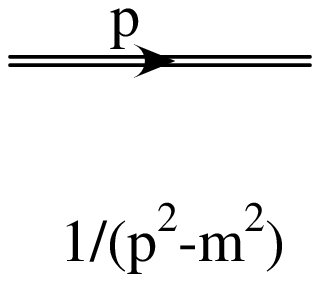}
  \quad
  \epsfxsize=2.12cm
  \epsffile[240 145 340 255]{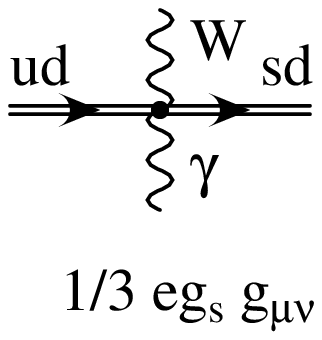}
 \end{center}
\vspace*{-.7cm}
\caption{Feynman rules for diquarks \label{frules}}
\end{figure}   
To perform the calculation one has to define Feynman rules for the 
radiation and the weak transition of the diquarks, which are simply 
obtained from the corresponding interactions of scalar particles in 
a straightforward way. Note that in order to ensure gauge invariance 
one also has to include a ``seagull'' term corresponding to a 
diquark-$W$-$\gamma$ coupling. The Feynman rules are given in fig.\ref{frules}
and these allow us a calculation of the hard part of the decay amplitude. 

The soft contributions are parametrized in terms of wave functions of 
the two constituents of the baryons. In the spirit of \cite{BrodskyLepage}
we neglect the transverse components of the momenta of the constituents. 
The $\Lambda_b$ is moving with the velocity $v$ and hence we write for the 
for the momentum $p$ of the $b$ quark $p=xm_b v$ and $k=(1-x)m_b v$ for 
the momentum of the $(ud)$ diquark. Similarly, we have $p'=y(m_b v-q)$
for the momentum of the $u$ quark in the $\Lambda$ and $k'=(1-y)(m_b v-q)$
for the momentum of its $(ds)$ diquark. The wave functions describing the 
soft part are $\Phi (x)$ and $\phi(y)$ for the $\Lambda_b$ and the $\Lambda$, 
which are
functions of the momentum fractions $x$ and $y$.
For the $\Lambda_b$ we assume that the $b$ quark 
carries almost all momentum, which means that $x$ fiffers from one only
by terms of the order $1/m_b$, such that $k$ is independent of $m_b$ in the 
heavy mass limit. Hence we use 
\begin{equation}
\Phi (x) = {1 \over 2 \sqrt{3}} f_B \delta(1-x) 
\end{equation}
for the wave functon of the $\Lambda_b$. 

For the $\Lambda$ all the quarks have comparable masses
and we use a simple  Ansatz such that the average of $y$ is $1/3$.
Motivated by the corresponding ansatz for the pion wave function we use 
\begin{equation}
\phi (y) =2 \sqrt{3}f_\pi  y(y-1)^2
\end{equation} 
where we have assumed that the analogue of the decay constant is 
simply $f_\pi$. 

Using this picture for the internal $W$ exchange it turns out that 
its contribution is much smaller than the long distance contribution 
of the vector dominance type. This is not a surprise, since the internal
$W$ exchange is on one hand suppressed by the small combination $V_{ub} 
V_{us}^*$ of CKM elements, on the other hand it contains two decay 
constants $f_\pi$ and $f_B$, yielding another suppression by a factor
$f_\pi f_B / m_b^2$

In total, in the way we have estimated the long distance contributions
they turn out to be small, such that the decay $\Lambda_b \to \Lambda \gamma$
is dominated by the short distance piece, which may be sensitive to new 
physics.  

Quantitatively, the long distance corrections are dominated by the 
vector-dominance contribution. Summing over the lowest six $\Psi^{(n)}$
we find that the long distance contributions may be absorbed into
a redefinition of the coupling constants $g_A$ and $g_V$ according
to
\beq \hat{g_V} = g_V + \rho \quad\mbox{and}\quad \hat{g_A} = g_A - \rho,
\quad\mbox{where}\quad \rho = 0.035 \label{rho}
\eeq

\section{Form Factors form the Semileptonic Decay 
         $\Lambda_c \to \Lambda \ell \bar{\nu}_\ell$}
Since we are dealing with an exclusive decay, we eventually need 
some input for the form factors in order to calculate numbers. 
However, not much is known about these functions, except that 
they have to obey the heavy quark symmetry implied by the
heavy-to-light transition we are considering. In particular, 
at least at maximum momentum transfer the form factors appearing
in $\Lambda_b \to \Lambda$ transition and $\Lambda_c \to \Lambda$ 
decays have to be the same, independent of the spin structure of 
the  currents involved, see (\ref{heavytolightdecay}). 

Again we shall assume heavy quark symmetry over all phase space
including $q^2 = 0$ in $\Lambda_b \to \Lambda$ and extract the 
relevant form factors from 
$\Lambda_c \to \Lambda \ell \bar{\nu}_\ell$, which is according to 
(\ref{heavytolightdecay}) given by the same two form factors as
 $\Lambda_b \to \Lambda \gamma$. For the semileptonic $\Lambda_c$ decay 
there is some experimental data \cite{cleo};
the total rate and the form factor ratio
$\left<F_2/F_1\right>$ has been measured, where the brackets indicate a 
particular average over phase space.

Nothing is known about the $q^2$ dependence of the form factors and 
at that point model assumptions enter the game. We shall start from 
a simple pole model for both form factors 
\begin{equation} 
F_{1/2}(q^2)=F^{\rm max}_{1/2} \cdot \left({M_{1/2}^2-m_Q^2
\over M_{1/2}^2-q^2}\right)^n, \quad n=1,2\, , \label{dipol} 
\end{equation}
where $M_{1/2}$ is the mass of the nearest resonance with the correct 
quantum numbers for $F_1$ and $F_2$ rspectively and $F^{\rm max}_{1/2}$
is a normalization factor. The form factors in (\ref{heavytolightdecay})
are given in terms of the energy $p.v$ of the outgoing light baryon 
which has a simple relation with the momentum transfer 
$q^2=m_Q^2+\ml^2-2m_Q (p.v)$. Inserting this into the denominator 
of our form factor parametrization we have 
\begin{equation}
M_{1/2}^2-q^2 = M_{1/2}^2+2m_Q (p.v) -m_Q^2-\ml^2 \approx
M_{1/2}^2+2m_Q (p.v)-m_Q^2
\end{equation}
where we have neglected the mass of the light baryon. 
The relevant resonances contain a heavy quark and some light 
degrees of freedom and hence the mass of the resonance is 
$M_{1/2} = m_Q + \Lambda_{1/2}$, where we shall in the following 
identify $\Lambda_1 = \Lambda_2 = \Lambda$. The parameter $\Lambda$ 
is small compared to the heavy quark mass and 
does not scale with the heavy mass; it remains a constant as 
$m_Q \to \infty$. Using this the denominator
of the pole fomula for the form factors becomes
\begin{equation}
M_{1/2}^2 -q^2 = (m_Q + \Lambda)^2 +2m_Q (p.v)-m_Q^2 
\approx 2m_Q (\Lambda + (p.v)) 
\end{equation}
For the $\Lambda_c \to \Lambda$ transition the relevant resonance 
for the vector current is the $D_s^*$ and since in the heavy mass limit 
$m_c \sim m_{\Lambda_c}$,
we have $\Lambda \approx m_{\Lambda_c}-m_{D_s^*} = 175$ MeV, 
while for the bottom decay $\Lambda_b \to \Lambda$ we have the $B_s^*$ 
in the vector channel, and hence here we obtain $m_{\Lambda_b}-m_{B_s^*}=
225$ MeV. Thus we have a fairly consistent picture and shall use 
$\Lambda = 200$ MeV, it turns out that the dependence of our results on 
$\Lambda$ is weak. 

Putting the pieces together we use as a parametrization of the 
form factors in the heavy mass limit
\begin{equation} \label{poleforms}
F_{1/2}(p.v)=N_{1/2} \cdot \left({\Lambda
\over \Lambda+p.v}\right)^n, \quad n=1,2\, , 
\end{equation}
whre we assume for simplicity the same $(p.v)$ dependence for both 
form factors. In this simple case the ratio
of the form factors $R = F_2 / F_1$
is independent of $(p.v)$ and we may use the 
value measured by CLEO without worrying about the phase space average
going into the measurement. We shall use \cite{cleo}
\begin{equation} 
R= F_2/F_1 = N_2/N_1 = -0.25 \pm 0.14 \pm 0.08 
\end{equation}
The large uncertainty in this value does not imply a large uncertainty in our 
results since any appearance of the form factor $F_2$ is suppressed by a 
factor $m_\Lambda / m_{\Lambda_Q}$. 

Using the measured branching ratio for $\Lambda_c \to \Lambda e^+ \nu$ 
and the lifetime of $\Lambda_c$ from \cite{pdg} one obtains 
the normalization $N_1$ 
\begin{equation} 
|N_1| = 7.06 \mbox{\quad (monopole)\quad \mbox{and}  \quad} 
|N_1| = 52.32 \mbox{\quad (dipole)} 
\end{equation}
Using the pole formulae (\ref{poleforms})
for the extrapolation to $q^2 = 0$ in the decay 
$\Lambda_b \to \Lambda \gamma$ we obtain for the form factors 
\begin{equation} 
F_1 (q^2 = 0) = 0.45 \mbox{\quad (monopole)\quad bzw. \quad}
F_1 (q^2 = 0)  = 0.22 \mbox{\quad (dipole)}. 
\end{equation}
This may be inserted into (\ref{gammashortdistance}) with the
corrections from (\ref{rho}) to obtain the total rate for the 
decay $\Lambda_b \to \Lambda \gamma$ in the standard model. 

In the heavy mass limit the mass of the $\Lambda_b$ is equal to
the $b$ quark mass and hence the total rate depends on $m_b^5$,
where some of the mass dependence comes from phase space 
(and is actually $m_{\Lambda_b}$) and some is due to the
$m_b$--dependence of the effective hamiltonian.
To get rid of this $m_b^5$--dependence we take as a reference
the inclusive semileptonic decays of the $B$ meson
and write:
\beq
{\Gamma(\Lambda_b \rightarrow \Lambda \gamma) \over 
\Gamma(B \rightarrow X_c l\nu)}
= {BR(\Lambda_b \rightarrow \Lambda \gamma) \over BR(B \rightarrow X_c l\nu)}
    \cdot {\Gamma_{\rm tot}(\Lambda_b) \over \Gamma_{\rm tot}(B)} 
\label{brlbtolg}
\eeq
The inclusive semileptonic decay rate of $B$ mesons depends also
on $m_b^5$ and is given by
\beq
\Gamma(B \rightarrow X_c l\nu)={m_b^5 G_F^2 V_{cb}^2 \over 192 \pi^3}
  \,\eta_{\rm QCD} \, r_c, \label{btoxlnu}
\eeq
where $\eta_{\rm QCD}= 0.94$ is a QCD correction
and $r_c=0.45$ is a phase space factor.

Using the lifetimes and the inclusive semileptonic branching fraction
$BR(B \rightarrow X_c l\nu)=(10.3 \pm 1.0) \%$
from \cite{pdg} we obtain
\beq
BR(\Lambda_b \rightarrow \Lambda \gamma)= 
    (1- 4.5) \cdot 10^{-5},
\eeq
where the lower and upper value corresponds to a dipole and monopole
$q^2$ evolution of the form factors, respectively. 

For values of $g_A$ and $g_V$ different from the standard model ones
one has to multiply the above equation by the factor $(g_A^2 + g_V^2)/2$, 
neglecting the strange quark mass and the long distance contributions.

\section{$\Lambda_b \to \Lambda \gamma$ as a Test of the Standard Model}
In this section we shall discuss the possibility of using the decay 
$\Lambda_b \to \Lambda \gamma$ as a test of the Standard model and to what 
extent it could be a possible window to new physics. 

As opposed to its mesonic counterpart, the decay $B \to K^* \gamma$, 
the baryonic rare decay $\Lambda_b \to \Lambda \gamma$ allows access
to the helicity structure of the short distance piece of the effective 
hamiltonian, since the helicity of the final state $\Lambda$ can be 
measured. The two relevant dimension-5 operators entering the short 
distance part are parametrized in 
terms of two coupling constants $C_7 g_A$ and $C_7 g_V$ and the 
polarization of the final state $\Lambda$ is sensitive to these couplings. 

The first piece of information is the total rate of 
$\Lambda_b \to \Lambda \gamma$ which contains (as does the mesonic decay 
$B \to K^* \gamma$) information on the combination 
$(g_V C_7)^2+ (g_A C_7)^2 $, 
while the polarization variable 
$\alpha$ defined in (\ref{alphadef}) contains information on 
$(C_7 g_A)/(C_7 g_V)$.   

As we have discussed above, our rough estimates of the long distance 
contributions indicate that they are very small and will not significantly 
influence the sensitivity of the $\alpha'$ measurement on the parameters of 
the short distance part of the effective hamiltonian. 

In the following we shall use the measurement of the inclusive rate 
$B \to X_s \gamma$ to fix $(g_V C_7)^2+ (g_A C_7)^2 $; 
in the $C_7 g_V$--$C_7 g_A$--plane a measurement of any total rate (meaning 
any of the processes $B \to X_s \gamma$, $B \to K^* \gamma$ or
$\Lambda_b \to \Lambda \gamma$) will correspond to a circle, since all 
these decay rates are proportional to $(g_V C_7)^2+ (g_A C_7)^2 $. 

If in future experiments a measurement of $\alpha'$ in 
$\Lambda_b \to \Lambda \gamma$ is performed one may use the relation 
between $\alpha'$ and $(C_7 g_A)/(C_7 g_V)$. Including our  
estimate of the long distance contributions, one obtains 
\begin{equation}
C_7 g_A = (C_7 g_V + C_7 \rho) 
          \left(\frac{0.353}{\alpha'} \pm 
          \sqrt{\left(\frac{0.353}{\alpha'}\right)^2 - 1} \right) 
                   + C_7 \rho
\end{equation}
which corresponds to straight lines in the $C_7 g_A$--$C_7 g_V$ plane. 
Neglecting the 
long distance contribution (i.e.\ $\rho = 0$)
these lines would go through the origin; including our estimate for 
these long distance contributions we also need an input for the 
Wilson coefficient  $C_7$, for which we shall use the standard model 
value as obtained form the NLLO calculation in \cite{chet}. 

\begin{figure}[ht]
\vspace{-.5cm}
 \begin{center}
  \leavevmode
  \epsfxsize=10cm
  \epsffile[70 160 540 630]{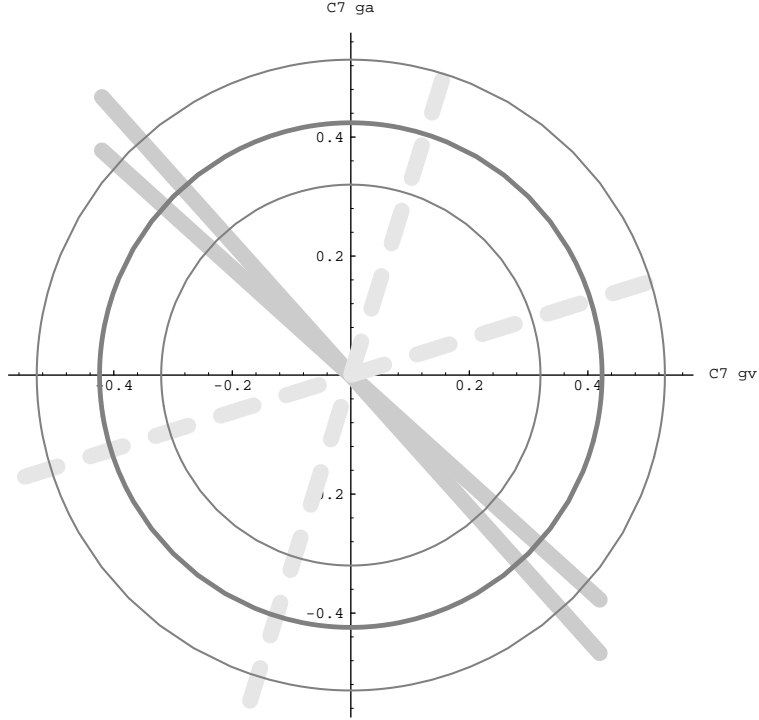}
 \end{center}
\vspace*{-.5cm}
\caption{The $C_7 g_A$--$C_7 g_V$ plane \label{gagvfig}}
\end{figure}
In fig.\ref{gagvfig} we plot the $C_7 g_A$--$C_7 g_V$ plane. 
The central circle corresponds to the central value of the measurement 
of $B \to X_s \gamma$ and the thin circles indicate the experimental 
uncertainty. Assuming the standard model values for $C_7 g_A$ and 
$C_7 g_V$ we have also plotted the corresponding lines. The width of 
these lines is given by our estimate for the long distance 
contributions; we have taken this estimate as an additional 
uncertainty meaning that the width of these lines is $2\rho C_7$. 

A measurement of $\alpha'$ yields in general two lines which have 
in total four intersections with the circle. The two intersections
of each line correspond to a sign interchange $g_A \to -g_A$ 
and at the same time $g_V \to -g_V$ which is an unobservable phase. 
The remaining ambiguity corresponds to the interchange 
$g_A \to g_V$ and $g_V \to g_A$, since the polarization variable 
and the total rate are symmetric functions of $g_A$ and $g_V$. 
Graphically this means that the two lines for a measured value of 
$\alpha '$ are mirror images of each other with respect to the lines 
$g_A = |g_V|$. In order to resolve this ambiguity additional 
measurements would be necessary. 

In the SM $|g_A| \approx |g_V|$ ($\alpha ' = -0.351$)
up to corrections from the not-vanishing $s$ quark mass,
therefore the two solid lines almost coincide for the SM value of
$\alpha'$. We also have plotted two dashed lines for a hypothetical 
measurement of $\alpha ' =0.2$.

\section{Conclusions}
With the advent of the second generation $b$ physics experiments,
bottom baryons will provide a wealth of additional information on 
$b$ quarks. In particular, specific aspects of FCNC decays can be 
tested which are not accessible in $B$ mesons decays. 

As its mesonic counterpart, $\Lambda_b \to \Lambda \gamma$ is 
dominated by the short distance part of the effective hamiltonian,
although there are additional long distance effects such as 
internal $W$ exchange. We have estimated the long distance effects 
using standard methods and found a negligibly small contribution. 

Thus one expects a good sensitivity to test the short distance 
part, the relevant piece of which is given in terms of two operators, 
which differ by the handedness of the $b$ quark. It is a particular
property of the decays of the $\Lambda_b$ that they allow one to test 
the handedness of the effective interaction; 
this is impossible in the corresponding meson decays. 

Heavy quark symmetries imply relations between the form factors which 
are thought to hold best close to the point where the outgoing hadron 
is almost at rest. For the transition $\Lambda_b \to$ light 
spin-1/2 baryon there are only two independent form factors
due to heavy quark symmetry. Measuring these form factors in 
$\Lambda_c \to \Lambda \ell \bar{\nu}_\ell$ then allows us to predict 
other processes. Unfortunately, in the process 
$\Lambda_b \to \Lambda \gamma$ we are considering the end of the 
$q^2$ spectrum, where the $\Lambda$ receives a large recoil of the 
order of the $b$ quark mass, and hence heavy quark symmetries might 
fail. We still used the relations implied by them, which in the 
worst case is simply a model assumption. 

Using the input for the form factors obtained from 
$\Lambda_c \to \Lambda \ell \bar{\nu}_\ell$ we made predictions for
the decay $\Lambda_b \to \Lambda \gamma$ within the standard model 
and beyond. It turns out that a measurement of the polarization of 
the final state $\Lambda$ is quite sensitive to the handedness 
of the underlying effective short distance hamiltonian, up to 
a twofold ambiguity, which cannot be resolved by a measurement 
of the rate and the polarization alone. 

With a branching ratio for $\Lambda_b \to \Lambda \gamma$ of the 
order $10^{-5}$ one needs $10^8$ $b$ quarks to have about one hundred 
events, without applying cuts for efficiencies. Clearly this will 
be feasible at dedicated $b$ physics experiments at colliders 
such as the one at Tevatron or LHC, and possibly also at fixed 
target experiments like HERA-B.

\section*{Acknowledgments}  
T.M. thanks A. Ali and A. Schenk for useful conversations. This 
work was supported by Deutsche Forschungsgemeinschaft.

\end{document}